# Surface residues dynamically organize water bridges to enhance electron transfer between proteins

Aurélien de la Lande,[1,2,‡] Nathan S. Babcock,[3,4] Jan Řezáč,[1,2,ə]
Barry C. Sanders,[3,4] Dennis R. Salahub[1,2,5,§]

[1] Department of Chemistry
[2] Institute for Biocomplexity and Informatics
[3] Department of Physics and Astronomy
[4] Institute for Quantum Information Science
[5] Institute for Sustainable Energy, Environment and Economy
University of Calgary, 2500 University Drive N. W., Calgary, Alberta, Canada, T2N 1N4

‡ *Present address*: Laboratoire de Chimie Physique—Centre National de la RechercheScientifique - Unité Mixte de Recherche 8000, Université Paris-Sud 11, Bâtiment 349, Campus d'Orsay. 15, Rue Georges Clémenceau, 91405 Orsay Cedex, France.
ə *Present address*: Institute of Organic Chemistry and Biochemistry, Academy of Sciences of the Czech Republic and Center for Biomolecules and Complex Molecular Systems, Flemingovo náměstí 2, 166 10 Prague 6, Czech Republic
§ To whom correspondence should be addressed: dsalahub@ucalgary.ca

**Abstract**

Cellular energy production depends on electron transfer (ET) between proteins. In this theoretical study, we investigate the impact of structural and conformational variations on the electronic coupling between the redox proteins methylamine dehydrogenase and amicyanin from *Paracoccus denitrificans*. We used molecular dynamics simulations to generate configurations over a duration of 40ns (sampled at 100fs intervals) in conjunction with an ET pathway analysis to estimate the ET coupling strength of each configuration. In the wild type complex, we find that the most frequently occurring molecular configurations afford superior electronic coupling due to the consistent presence of a water molecule hydrogen-bonded between the donor and acceptor sites. We attribute the persistence of this water bridge to a "molecular breakwater" composed of several hydrophobic residues surrounding the acceptor site. The breakwater supports the function of nearby solvent-organizing residues by limiting the exchange of water molecules between the sterically constrained ET region and the more turbulent surrounding bulk. When the breakwater is affected by a mutation, bulk solvent molecules disrupt the water bridge, resulting in reduced electronic coupling that is consistent with recent experimental findings. Our analysis suggests that, in addition to enabling the association and docking of the proteins, surface residues stabilize and control interprotein solvent dynamics in a concerted way.

**Introduction**

The electron transport chain is the cornerstone of biological energy transduction. All known life-forms use membrane-bound chains of redox proteins to convert energy from food or sunlight into chemical energy stored in adenosine triphosphate (1). Biological electron transfer (ET) often occurs over long distances (>1nm) between protein-encapsulated redox cofactors separated by intervening protein or solvent molecules. Over the last two decades, there has been increasing interest in water as an "active constituent in cellular biology" (2). Today there is a growing body of evidence suggesting that water plays an important role mediating long-range ET and that conformational fluctuations are critical to protein-solvent interactions at ET interfaces (3,4,5,6,7,8,9,10). Notably, previous authors have suggested that "ordered water molecules in the protein-protein interface may considerably influence electronic coupling between redox centers" (4) and that "water may be a particularly strong tunneling mediator when it occupies a sterically constrained space between redox cofactors with strong organizing forces that favor constructively interfering coupling pathways" (7).

In general, however, the degree of sophistication of solvent-organizing effects at aqueous ET interfaces remains unknown. In this study, we show that a pair of solvent-organizing residues in direct contact with a bridging water molecule may be aided by surrounding residues that help stabilize local solvent dynamics by mediating contact with the bulk. We predict that surface residues at protein-protein interfaces can act collectively to organize and stabilize solvent structures and dynamics during long-range ET.

The timely passage of electrons from protein to protein is crucial for proper metabolic regulation (11) and relies on all the physical and chemical phenomena (*i.e.*, diffusion, protein docking, and ET reaction steps) that participate in overall electron transmission. Here we investigate the redox reaction between methylamine dehydrogenase (MADH) and amicyanin taken from *Paracoccus denitrificans*. This redox pair is representative of a broad class of interprotein ET reactions involving blue copper proteins (12). Under methylotrophic growth conditions, MADH supplies electrons to amicyanin, a blue copper protein that in turn shuttles electrons to various c-type cytochromes (13). *In vitro*, the transfer occurs between the reduced tryptophan tryptophylquinone (TTQ) group on the MADH β subunit and a cupric complex buried just under the amicyanin surface (Fig. 1). The oxidation of MADH by amicyanin is a "true" ET reaction, limited by the ET reaction rate $k_{ET}$ (14). Ma *et al.* recently reported a series of site-directed mutations performed on the amicyanin methionine 51 residue by alanine, lysine and leucine (Fig. 1) (15). Kinetic measurements revealed nearly a tenfold decrease in $k_{ET}$ without significant changes to the proteins' overall structural, binding, or redox potentials. Ma *et al.* concluded that "surface residues of redox proteins may not only dictate specificity for their redox protein partners but also be critical to optimize the orientations of the redox centers and intervening media within the protein complex for the ET event" (15).

To analyze an interprotein ET reaction like the reduction of amicyanin by MADH, it is necessary to consider large ensembles of protein-protein complex configurations that contribute to the average ET rate (16). For a "true" ET reaction, the ET rate $k_{ET}$ can be estimated by nonadiabatic Marcus-Hush-Levich theory (17,18),

$$k_{ET} = \frac{2\pi}{\hbar} \frac{1}{\sqrt{4\pi\lambda k_B T}} |T_{DA}|^2 \exp\left(\frac{-\Delta G^\ddagger}{k_B T}\right) = \frac{2\pi}{\hbar} \frac{1}{\sqrt{4\pi\lambda k_B T}} |T_{DA}|^2 \exp\left(\frac{-(\Delta G^\circ + \lambda)^2}{4\pi k_B T}\right), \quad (1)$$

where $\hbar$ is the reduced Planck's constant, $k_B$ is Boltzmann's constant, $T$ is the temperature, $\Delta G^\ddagger$ is the Gibbs free energy of activation, $\Delta G^\circ$ is the Gibbs free energy of reaction, $\lambda$ is the reorganization energy, and $T_{DA}$ is the superexchange matrix element which couples the donor and acceptor electronic states quantum mechanically.

The experimental trends reported by Ma *et al.* strongly suggest the existence of a correlation between protein motion and ET activity in the MADH–amicyanin complex. While most numerical studies have addressed nanosecond timescales or shorter, we investigated longer timescales (40ns) that are less well understood, thereby obtaining statistics for vibrational modes spanning several temporal orders of magnitude. In addition to the wild type complex, we consider four amicyanin mutants: M51A, M51L, M51K and M51C. The first three mutations correspond to those reported by Ma *et al.* (15), whereas M51C was added to investigate the impact of a thiol replacing the original thioether. We employed Molecular Dynamics (MD) simulations to generate the configurations within the transient ET complex along with an ET pathway analysis to characterize each configuration's intrinsic ET activity (see Methods section). This computationally intensive investigation has allowed us to identify important solvent-stabilizing functions of interprotein surface residues. Because of this solvent-stabilizing effect, the most ET-active wild type conformations are also the most statistically favored ones, an effect that is lost in the mutant complexes.

**Results**

We obtained 400 000 molecular configurations from 40ns of Molecular Dynamics simulations of the wild type complex and each mutant. We employed the semi-empirical pathway model originally developed by Beratan, Onuchic and Hopfield (19) to determine the tunneling pathway with the largest electronic coupling matrix element $T_{DA}$ for each configuration. Although this model does not provide an absolute value for the superexchange coupling matrix element $T_{DA}$ and does not account for interferences between multiple pathways, it can be used to estimate the total electronic coupling decay factor $\varepsilon_{tot}$, where $T_{DA} \approx H_{DA} \cdot \varepsilon_{tot}$ and $H_{DA}$ is the theoretical "close contact" coupling matrix element (20,21). Despite its simplicity, the pathway model has previously demonstrated excellent predictive power when comparing different molecular configurations (8,22,23). We recorded the pathway with the largest decay factor $\varepsilon_{tot}$ for each configuration and labeled it by the surface residue through which the electron exits the MADH. We found that in the vast majority of pathways (>99%), the electron leaves the MADH through one of Ser β 56, Trp β 57, Val β 58, or Trp β 108 before tunneling through one or more water molecules and entering the amicyanin through His 95. Adopting the terminology already in use (21), it is convenient to define four distinct collections of similar pathways, or "pathway tubes," labeled by the letters A (Ser β 56), B (Trp β 57), C (Val β 58), and D (Trp β 108) (Figs. 2 and S1). All the remaining excess pathways (labeled E) afford comparatively weak electronic coupling (Table S1).

Pathway tube A is of particular interest due to its large coupling strength and high frequency of occurrence (~60%, Table 1). For this reason, we further divided it into three subcategories: $A_1$ represents a single pathway with a hydrogen-bonded bridge from the Ser β 56 O carbonyl oxygen through a single water molecule to the His 95 HE2 proton; $A_2$ represents all the remaining completely hydrogen-bonded water bridges between Ser β 56 and His 95; and $A_3$ represents all the partially broken (*i.e.*, van der Waals coupled) pathways from Ser β 56 to His 95. In a previous pathway analysis performed on the crystal structure of the MADH−amicyanin dimer (24), Brook *et al.* found that the strongest pathway required a through-vacuum jump from MADH Trp β 108 to amicyanin Pro 94 (25). In contrast, in our solvated system we find that pathways involving a direct jump from Trp 108 to Pro 94 make up less than 0.01% of our data and are on average one tenth as strong as the completely hydrogen-bonded $A_1$ pathway. In another analysis performed on the MADH−amicyanin−cytochrome ternary crystal structure, Chen *et al.* concluded that the strongest pathway involved a trapped interfacial water molecule, even though it's efficiency depended "critically on the presence of the water molecule which may not always be occupied" (26). Chen *et al.*'s water-mediated pathway is identifiable as our $A_1$ pathway, assuming a hydrogen-bonded arrangement for the hydrogen atoms that were not resolved. Thus, our computational study strongly supports Chen *et al.*'s water-bridge hypothesis and moreover stresses the importance of the dynamical behavior of the ET pathways; very frequent switches between the pathways are obtained in the course of the simulations (Fig. S2) but pathway $A_1$ was favored *only* in the wild type complex.

Configurations associated with pathway tube $A_1$ depend critically on the probability $P_{hb}$ of a water molecule forming two simultaneous hydrogen bonds with atoms Ser β 56 O and His 95 HE2. The wild type complex's statistical affinity for pathway $A_1$ depends on this high probability ($P_{hb} > 50\%$) compared with those of the mutants ($P_{hb} < 20\%$) (Table 1). In turn, the presence of this water bridge is linked to the discrepancy in the average number of water molecules found at the interface between the proteins. The consistent presence of the water molecule joining Ser β 56 and His 95 in the wild type, its corresponding absence in the mutants, and the resulting reduction in the mutant coupling strength indicate that solvent organization is vital to this reaction. Any destabilization of the $A_1$ water bridge results in a statistical shift towards less efficient pathways (Table 1).

The $A_1$ pathway is disrupted when other water molecules jostle or compete with the bridging water molecule (Fig. 2 $A_1$). To determine the impact of surrounding water molecules on the $A_1$ bridge, we computed the number of water molecules within the "ET region", which we define to be a sphere of radius R centered between the MADH Ser β 56 O and amicyanin His 95 HE2 atoms. Even for large radii (R = 5Å), an average of only 2.5 water molecules are present within the wild type ET region, compared with mutant averages ranging from 4.7 to 6.3 (Fig. 3). Comparatively few water molecules are exchanged between the wild type ET region and the surrounding bath. The mutant ET regions are much more turbulent, as evidenced by the larger number of water molecules present, the lower probability of hydrogen-bonded bridge formation $P_{hb}$, and the higher rate of turnover τ between the individual water molecules involved in forming the $A_1$ bridge (Table 1). These results indicate that the dynamical organization of the intervening solvent is crucial to the formation of the most efficient ET configurations.

A comparison of mutant-to-wild type ratios of the calculated decay factors $r_\varepsilon^{mut} = \langle \varepsilon_{tot}^2 \rangle^{mut} / \langle \varepsilon_{tot}^2 \rangle^{wt}$ and the experimental rates $r_k^{mut} = k_{ET}^{mut} / k_{ET}^{wt}$ provides insight about the impact of the Met 51 mutation on experimental rates through modifications of the electronic coupling term. Our results are in overall qualitative agreement with experiment, as the decreases in the mutant decay factors are of comparable size to the experimentally observed decreases in the ET rate constants (Table 1). There is, however, a discrepancy between the ordering of the experimentally determined mutant rates $(1 > r_k^{M51L} > r_k^{M51K} > r_k^{M51A})$ and the pathway model decay factors $(1 > r_\varepsilon^{M51A} > r_\varepsilon^{M51K} > r_\varepsilon^{M51L})$, discussed below. We find that the packing density model (27) produces decay factors similar to those of the pathway model for the M51L and M51K mutants, but predicts the M51A and M51C mutants to be at least as kinetically competent as the wild type.

**Discussion**

It is remarkable that a single strongly-coupled pathway should dominate the thermal statistics of the wild type complex alone (>50%, Fig. 3a). There is no *a priori* correlation between the binding affinity of the complex and ET activity: Maxwell-Boltzmann statistics govern the probability of occurrence of a given configuration, whereas the non-adiabatic Marcus expression (eq. 1) independently determines the ET rate constant $k_{ET}$ for that configuration. It is apparent that the protein structure at the wild type interface is specifically suited to favor conformations most amenable to ET. On the other hand, configurations associated with lower efficiency pathways become more statistically prominent in the mutant distributions, thereby decreasing the average decay factor $\langle \varepsilon_{tot} \rangle$ (Table 1). This statistical change also carries implications for the overall reaction kinetics.

In order to relate solvent dynamics directly to the mutation, it is necessary to consider interactions between the amicyanin 51 residue and other protein or solvent molecules. Previously, when performing a P52G mutation upon amicyanin, *Ma et al.* attributed the resulting reduction in $k_{ET}$ to a loss of interactions between the amicyanin Met 51 residue and the MADH Val 58 (28). This conclusion is compatible with our simulation results, which show that conformational variations in the amicyanin M51K and M51L residues allow an increased number of water molecules to "sneak" into the ET region (Fig. S3). Intuitively, one expects the replacement of Met 51 by a smaller (alanine) or hydrophilic (lysine) residue to allow more water molecules to enter the ET region. The cases of cysteine and leucine are less straightforward to analyze, as both residues are hydrophobic and similar in size to methionine. Many complex interactions influence the dynamics of these residues, and a future analysis will have to examine a variety of chemical effects (*e.g.,* methionine is a Lewis base whereas leucine is more acidic). In this regard, our study highlights the importance of subtle interprotein surface dynamics to the formation of efficient ET pathways.

Ma *et al.* used the "*true, gated, and coupled ET*" (17) framework to rationalize the decrease in the ET rates in terms of protein motion at the interface. Based on large increases in the inferred values of $T_{DA}$ and $\lambda$, as well as observed changes to the rate-limiting reaction kinetics for the N-quinol TTQ form of MADH (15,28), Ma *et al.* inferred that the rate constants $k_{ET}$ measured for the M51A and M51K mutants were not those of the "true" ET reaction, as is the case for the wild type system. Rather, they proposed that the M51A and M51K reactions were "gated" by an unidentified, separate, slower pre-ET step *x* that imposed its rate $k_x$ over that of the actual ET event (15). On the other hand, experimental data for the M51L mutant is similar to that of the true wild type reaction and is not consistent with conformationally gated ET for which $k_{ET} > k_x$. For the M51L mutant, Ma *et al.* concluded that either $k_x \sim k_{ET}$ or that the ET reaction is kinetically "coupled" to a rapid but unfavorable conformational rearrangement with equilibrium constant $K_x$, so that the observed rate is actually $k_{ET} \times K_x$ (15). This kinetically coupled picture (29) is compatible with the "dynamic docking" framework in which "a large ensemble of weakly bound protein-protein configurations contribute to binding, but only a few are reactive" (30). It is not clear why the mutation of the MADH Met 51 residue would lead to gated ET ($k_{ET} > k_x$) in the M51A and M51K mutants, but coupled ET ($k_{ET} < k_x$) in the M51L mutant.

Our numerical analysis is consistent with the viewpoint that ET is modulated by rapid interconversion within an ensemble of configurations of varying ET reactivity. The configurations produced by our simulations exhibit a continuum of ET affinities, whereas the kinetically coupled and dynamic docking models assume a simple active/inactive model of ET activity (29,30). This active/inactive dichotomy fails to capture the variation in intermediate coupling strengths revealed by our pathway analysis (Fig. 3). ET rate reductions comparable to the experimental ones are obtained by summing the contributions to $T_{DA}$ arising from the various accessible molecular configurations within the transient ET complex, without assuming a distinct pre-organization step. Because each configuration is associated with an intrinsic ET coupling strength, it is enough to modulate the ET rate simply by modifying each configuration's respective statistical weight. We propose the hypothesis that the increased amount of water at the ET interface dynamically modulates the ET rate in the mutants, akin to kinetically coupled ET as described above.

Further work will be required to reproduce the exact experimental trend in the mutant rate constants (Table 1). Variations in relevant parameters like $\Delta G°$ and $\lambda$ contribute to the experimental rate $k_{ET}$, but given that the wild type ET rate is $k_{ET} = 10s^{-1}$, 40ns of MD simulation may not fully account for these parameters. Furthermore, although the use of Langevin dynamics improves the sampling of the configuration space, the artificial noise inherent to this method can also become a source of error. The pathway model itself is limited by its inability to account for complex-valued interferences between tunneling pathways. It successfully estimates the electronic coupling for molecular configurations with a single dominant tunneling-pathway or a few constructively-interfering pathways, but its accuracy is limited for configurations with multiple destructively-interfering pathways (7). Because pathway "tube" $A_1$ represents only one strongly-coupled pathway and because very few water molecules are present at the wild type interface, the pathway model is expected to provide a good coupling estimate for the statistically-favored wild type $A_1$ configurations. The increased number of water molecules at the mutant interfaces makes inter- and intra-tube destructive interference more likely in the mutant complexes, and the pathway model may over-estimate the coupling strengths for these configurations. The question of multiple interferences accentuates the potential importance of solvent control to create one dominant strongly-coupled pathway at the protein interface.

**Conclusions**

Earlier studies on both inter- and intraprotein ET revealed the possibility of water-mediated ET pathways in biological ET (6,31), as well as the specific role of protein residues stabilizing well-defined ET pathways (8,32). Our study, however, provides the first evidence that several protein surface residues can act together in concert to organize bridging water molecules, enhancing electron transfer between proteins. Our numerical simulations indicate that MADH Ser β 56 and amicyanin His 95 work together to form a solvent-linked bridge between donor and acceptor, while the surrounding hydrophobic residues act as a "molecular breakwater" to support the stability of this bridge (Fig. 1b). Comparisons of the solvent organization in the wild type and mutant complexes show that the amicyanin Met 51 residue plays an essential role, repelling bulk water molecules from the ET region (Fig. 4). Any modification of the steric or electrostatic interactions at the Met 51 site—by either replacement (15) or repositioning (28)—may disrupt this solvent-repelling mechanism. In this respect, we believe that site-directed mutagenesis studies of the nearby amicyanin Met 28 and Met 71 residues (Fig. 1) would also be of great interest. If Met 28 and Met 71 function in the same manner as Met 51, mutations to these residues will produce reductions in $k_{ET}$ similar to those found for Met 51.

More generally, our proposed solvent repelling mechanism depends on a patch of hydrophobic surface residues surrounding the acceptor site, a characteristic shared by other blue copper proteins (12). So far, this surface characteristic was believed to ensure a weak binding affinity of the redox partners, allowing fast association and dissociation processes. Our study reveals another possible role for a blue copper protein's hydrophobic surface (Fig. 1b). It may enhance the ET activity of the redox complex, controlling solvent dynamics to significantly improve the strength and stability of water-mediated ET pathways.

**Methods**

*Molecular Dynamics Simulations*. We carried out Molecular Mechanics computations using the CHARMM 33a package (33). We selected the ternary MADH–amicyanin–cytochrome-c551i complex resolved to 1.9 Å by X-ray crystallography (Protein Data Bank entry 2GC4). This is a reasonable starting structure for simulations since crystalline MADH has been demonstrated to be catalytically competent to transfer electrons to amicyanin (34,35). After deleting the cytochrome-c551i from the ternary complex, hydrogen atoms were added with the HBUILD routine (as implemented in CHARMM) and the proteins were solvated in a TIP3P (36) water box of dimensions 115×80×80 Å$^3$. Approximately 40 Na$^+$ ions were added to ensure electrical neutrality (depending on the mutant). Histidine residues, including His 53 and His 95 were mono-protonated consistent with the experimental pH of 7.5 (15). The mutant complexes were generated from this structure *in silico* using Molden (37). The amicyanin cupric center was treated using the Force Field parameters developed by P. Comba *et al.* for blue copper proteins (38). The Lennard-Jones parameters for the copper ion were ε = 0.05 kJ/mol and σ = 2.13 Å (39). The wild type and mutant structures were first geometrically optimized by 500 steps of steepest descent algorithm and subsequent 1500 steps of Adopted Basis Newton-Raphson optimizer. This was followed by 1ns of Langevin dynamics to ensure equilibration and a further 40ns from which configurations were sampled every 100fs. The Shake algorithm was employed to constrain hydrogenated bonds at their equilibrium bond lengths. A friction coefficient of 15 ps$^{-1}$ and a bath temperature of 298K were used to propagate the equations of motion within the Langevin approach. Periodic boundary conditions were applied to simulate a continuous medium. Finally, a shift function was used to compute electrostatic interactions between distant pairs of atoms, with a 12 Å cut-off. A switch function was applied for van der Waals interactions (starting from 10 Å and set to zero at 12 Å). This is the recommended (default) scheme in CHARMM to compute non-bonded terms.

*Choice of Donor and Acceptor.* We defined the donor based on the Density Functional Theory (DFT) Highest Occupied Molecular Orbital (HOMO) of the tryptophan tryptophylquinol (TTQ) cofactor. For these computations we used the deMon2k code (40) with the Perdew–Burke–Ernzerhof functional (41) and the DZVP-GGA basis sets. The catecholate ring represents 63% of the donor molecular orbital, whereas the full MADH Trp β 57 aromatic ring represents almost 73%. There is very little orbital delocalization onto the MADH Trp β 108 residue (less than 15% spread over its aromatic ring) and as such it cannot be considered part of the donor group. We therefore restrict our definition of the donor to the Trp β 57 catecholate ring, assigning a decay factor of 1 between the Trp β 57 atoms. Consequently, the best pathway for a given configuration does not depend on the choice of the starting atom within the MADH Trp β 57 catecholate ring. We note that our DFT-based definition of the donor orbital is different from the one chosen in a previous pathway analysis where the electron density was assumed to be delocalized across both cycles of the TTQ group and the Trp β 108 residue was therefore taken as part of the donor group (25). The copper atom was taken as the acceptor since the beta Lowest Unoccupied Molecular Orbital (LUMO) essentially consists of the copper $d_{xy}$ orbital (some contributions are found on the Cys 92 residue but do not extend further than the sulphur atom 3p orbital).

*ET analysis.* We chose the empirical pathway model originally developed by Beratan *et al.* (20) to estimate $\varepsilon_{tot}$ for the huge number of sampled molecular configurations. The pathway model allowed us to classify the configurations in terms of distinct geometric motifs, directly relating conformational fluctuations to variations in the coupling strength. The pathway model assumes that the electron can tunnel from atom to atom along a given pathway, each interatomic step *i* contributing a coupling decay factor $\varepsilon_i$. Individual covalently-bonded, hydrogen-bonded, and through-vacuum decay factors (denoted $\varepsilon_c$, $\varepsilon_{hb}$, $\varepsilon_v$, respectively) were calculated based on semi-empirical formulae (eqs. 3-6). $T_{DA}$ is the product of the first order close-contact matrix coupling element $H_{DA}$ and the total semi-empirical decay factor $\varepsilon_{tot}$ (eq. 3), where $\varepsilon_{tot}$ is the product of N individual decay factors ($N = N_c + N_{hb} + N_v$, respectively). To improve the accuracy of $\varepsilon_{tot}$, we used refined parameters derived recently from constrained Density Functional Theory (DFT) (42,43) for the $\varepsilon_{hb}$ term which depends on the hydrogen-bond angle $\phi$ and the atom-to-atom distance R. We employed Dijkstra's algorithm (44) to find the pathway with the largest coupling for each configuration. To make the search tractable, each protein complex was pruned to about 300 atoms at the ET interface belonging to the following residues: amicyanin Met 28, Met 51, Pro 52, His 53, Met 71, Cys 92, Pro 94, His 95, Met 98, Cu(II); MADH Ala β 55, Ser β 56, Trp β 57, Val β 58, Pro β 100, Glu β 101, Trp β 108; and all water molecules within 7 Å of the amicyanin 95 HE2 or MADH 108 CD2 atoms.

$$T_{DA} = H_{DA} \cdot \varepsilon_{tot} \quad (2)$$

$$\varepsilon_{tot} = \prod_{i=1}^{N_c} \varepsilon_c^i \cdot \prod_{j=1}^{N_{hb}} \varepsilon_{hb}^j \cdot \prod_{k=1}^{N_v} \varepsilon_v^k \quad \text{[pathway model]} \quad (3)$$

$$\varepsilon_c = 0.6 \quad (4)$$

$$\varepsilon_{hb} = 0.36 \cdot e^{-0.64(R-2.01)} \cdot e^{-2.23(\cos\phi+1)} \cdot e^{-1.83(R-2.01)(\cos\phi+1)} \quad (5)$$

$$\varepsilon_v = 0.6 \cdot e^{-1.7(R-1.4)} \quad (6)$$

For comparison with the pathway model, the packing density (27) approach was also tested. In this case, $\varepsilon_{tot}$ is written as a product of two exponential decay factors that involve the fraction of filled space (*i.e.*, space within the atoms' van der Waals radii), the complementary fraction of vacuum space, and the donor-acceptor separation $R_{DA}$ (eq. 8). The associated decay factor parameters $\beta_{fill}$ (0.45 Å$^{-1}$) and $\beta_{vac}$ (1.4 Å$^{-1}$) were taken from Page *et al.* (27). To evaluate the fraction of filled space $f_{fill}$ we defined 200 points regularly spaced along the donor-acceptor axis and determined for each of them the presence of any surrounding atom within their van der Waals radius.

$$\varepsilon_{tot} = e^{-f_{fill}\beta_{fill}R_{DA}} \cdot e^{-(1-f_{fill})\beta_{vac}R_{DA}} \quad \text{[packing density model]} \quad (7)$$

*Figures.* Molecular graphics were prepared with VMD (version 1.8.6) (45).

**Acknowledgements**

The authors wish to thank Sergei Noskov (Univ. Calgary, Canada) and Pascal Permot (Univ. Paris-Sud 11 – CNRS, France) for helpful discussions. This work was supported by the Natural Sciences and Engineering Research Council of Canada (NSERC), the Canadian Institute for Advanced Research (CIFAR), the Informatics Circle of Research Excellence (iCORE), and the Alberta Ingenuity Fund (AIF). Computational resources were provided by WestGrid. B.C.S. is a CIFAR Fellow.

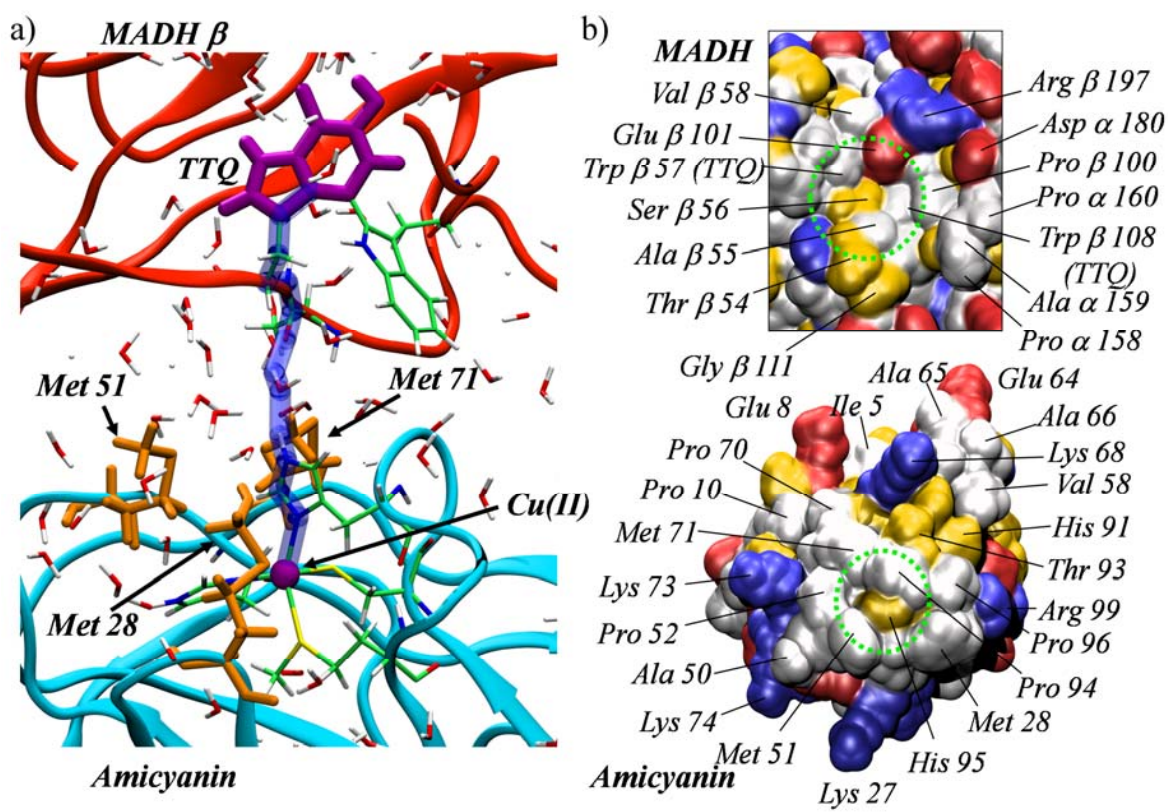

Figure 1: a) Solvated amicyanin (blue) in contact with MADH subunit β (red). Residues of interest to ET are represented as liquorice. Redox cofactors are shown in purple and surface methionine residues in orange. The dominant pathway $A_1$ is shown as a transparent blue tube connecting the redox cofactors. b) Direct ("head on") view of interfacial residues represented by their chemical nature: hydrophobic (white), hydrophilic (yellow), positively charged acidic (blue) and negatively charged basic (red). The dotted green circles indicate the ET region on the surface of each protein. The "molecular breakwater" is visible as a white ring of hydrophobic residues surrounding the amicyanin His 95.

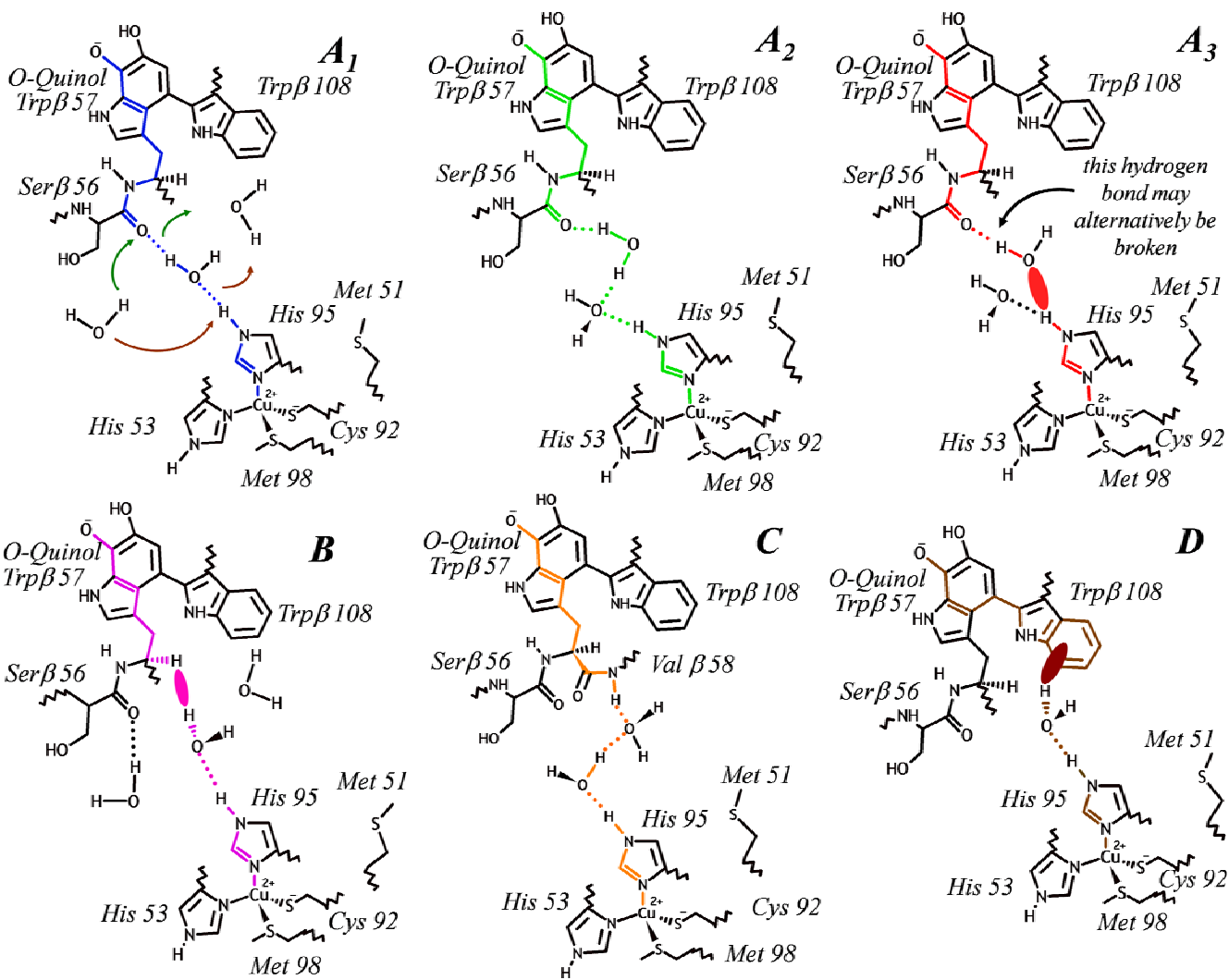

Figure 2: Representative pathways for each pathway tube in colors corresponding to those in Fig. 3. Hydrogen bonds are represented by dotted lines and through-space jumps by solid ovals. The arrows on $A_1$ illustrate perturbations to the hydrogen bond network caused by another nearby water molecule.

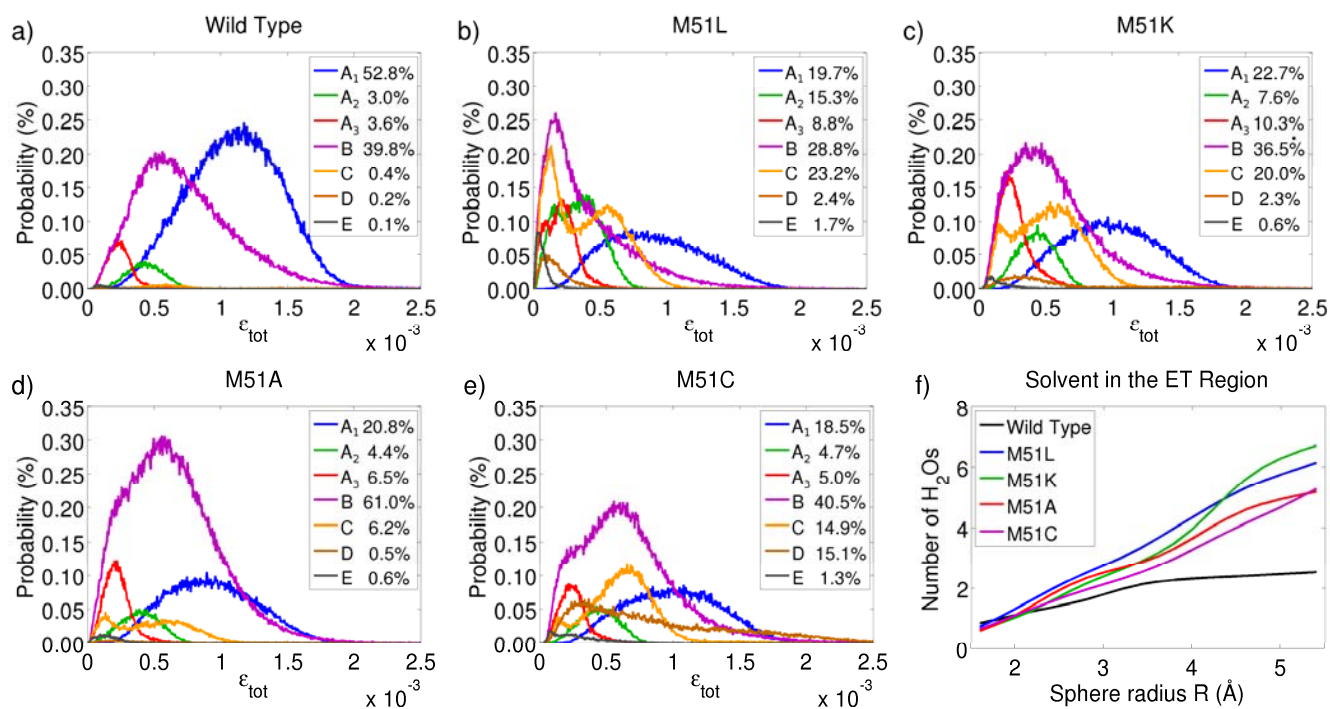

Figure 3: Normalized distributions of sampled pathway tubes (a-e) and the average number of water molecules (f) within the ET region defined by a sphere of radius R (sphere shown in Fig. 4).

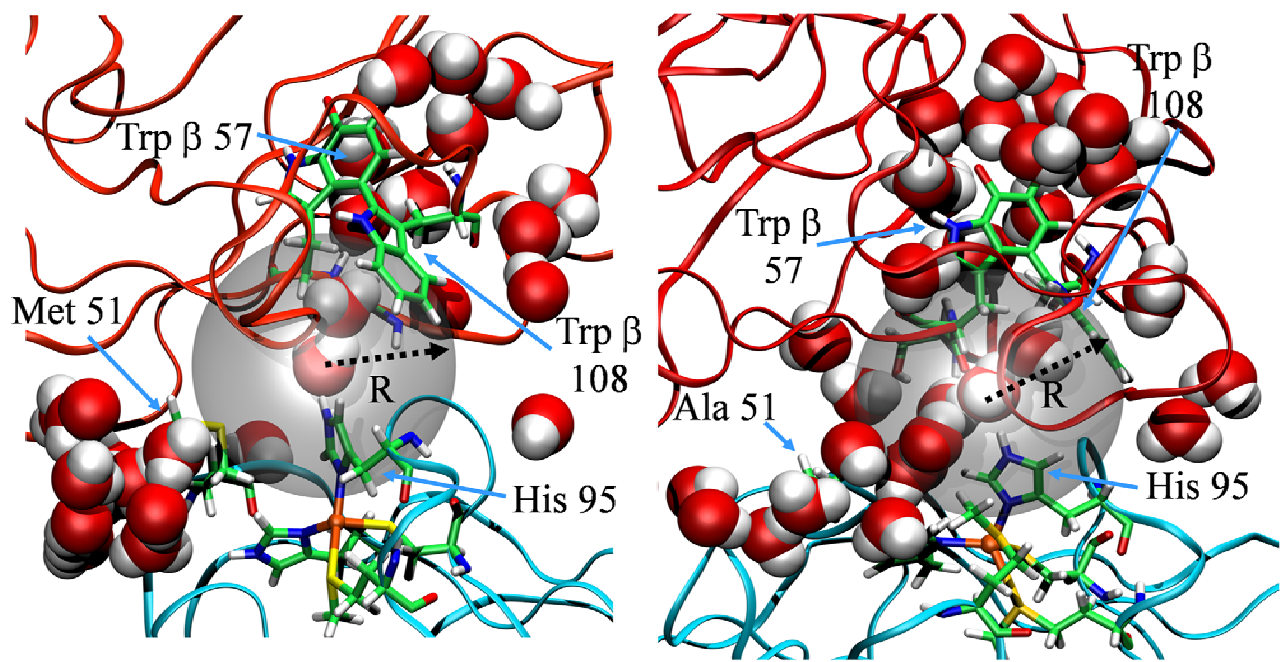

Figure 4: Representative snapshots of the wild type (left) and M51A (right) interfaces, revealing the breach in the molecular breakwater due to the mutation (other mutants shown in Fig. S3). The grey sphere represents the ET region for radius R = 5.5 Å. Water molecules are shown in the van der Waals representation.

|  |  | wild type | M51L | M51K | M51A | M51C |
|---|---|---|---|---|---|---|
| $r_k^{mut}$ | (experiment) | 1.0 | 0.68 | 0.49 | 0.13 | — |
| $\langle \varepsilon_{tot} \rangle \times 10^3$ | (pathway analysis) | 0.90 ± 0.03 | 0.47 ± 0.03 | 0.61 ± 0.02 | 0.65 ± 0.02 | 0.73 ± 0.02 |
| $r_\varepsilon^{mut}$ | (pathway analysis) | 1.0 | 0.36 ± 0.04 | 0.52 ± 0.04 | 0.57 ± 0.04 | 0.76 ± 0.05 |
| $\langle \varepsilon_{tot} \rangle \times 10^3$ | (packing density) | 0.70 ± 0.03 | 0.42 ± 0.04 | 0.51 ± 0.03 | 0.62 ± 0.05 | 1.03 ± 0.05 |
| $r_\varepsilon^{mut}$ | (packing density) | 1.0 | 0.56 ± 0.09 | 0.76 ± 0.07 | 0.89 ± 0.15 | 2.29 ± 0.26 |
| $P_{hb}$ |  | 0.53 | 0.15 | 0.19 | 0.18 | 0.16 |
| $\tau$ (ns$^{-1}$) |  | 0.40 | 0.57 | 0.60 | 0.65 | 0.68 |

Table 1: Expectation values for $\langle \varepsilon_{tot} \rangle$ and the ratios $r_\varepsilon^{mut} = \langle \varepsilon_{tot}^2 \rangle^{mut} / \langle \varepsilon_{tot}^2 \rangle^{wt}$ and $r_k^{mut} = k_{ET}^{mut} / k_{ET}^{wt}$ obtained from packing density and pathway analysis of the configurations gathered along the molecular simulations. The uncertainties account for the sampling errors of the computational protocol (see Supporting Information). Experimental rates $k_{ET}$ were obtained from $k_3$ (at 30°C) in Table 3 of ref. (15) (M51C was not reported). $P_{hb}$ is the unit-normalized likelihood that a water molecule is simultaneously hydrogen bonded to both the MADH Ser β 56 O and amicyanin His 95 HE2 atoms during our simulations. The turnover $\tau$ of the bridging water molecule is defined as the number of different water molecules that participate in pathway $A_1$ divided by the length of the simulation in nanoseconds.